*Type of the Paper (Article, Review, Communication, etc.)*

# Autonomous Identity-Based Threat Segmentation in Zero Trust Architectures

Sina Ahmadi[1]

[1] The National Coalition of Independent Scholars (NCIS), sina0@acm.org

**Abstract:-** Zero Trust Architectures (ZTA) fundamentally redefine network security by adopting a "trust nothing, verify everything" approach that requires identity verification for all access. Conventional discrete access control measures have proven inadequate since they do not consider evolving user activities and contextual threats, leading to internal threats and enhanced attacks. This research applies the proposed AI-driven, autonomous, identity-based threat segmentation in ZTA, along with real-time identity analytics for fine-grained, real-time mechanisms. Some of the sharp practices include using the behavioral analytics approach to provide real-time risk scores, such as analyzing the patterns used for logging into the system, the access sought, and the resources used. Permissions are adjusted using machine learning models that take into account context-aware factors like geolocation, device type, and access time. Automated threat segmentation helps analysts identify multiple compromised identities in real-time, thus minimizing the likelihood of a breach advancing. The system's use cases are based on real scenarios; for example, insider threats in global offices demonstrate how compromised accounts are detected and locked. This work outlines measures to address privacy issues, false positives, and scalability concerns. This research enhances the security of other critical areas of computer systems by providing dynamic access governance, minimizing insider threats, and supporting dynamic policy enforcement while ensuring that the needed balance between security and user productivity remains a top priority. We prove via comparative analyses that the model is precise and scalable.

## 1. Introduction

*1.1. The Shift Towards Zero Trust Architectures*

Traditional approaches to cyber protection have been rendered ineffective due to new and complex cyber threats and the rapid expansion of digitization in business organizations [1]. Perimeter security models, which assume a clear divide between secure internal networks and dangerous external ones, have proven insufficient against advanced persistent threats (APTs), insiders, and remote access threats. This has led to the development of Zero Trust Architecture (ZTA), which operates on the fundamental principle of "never trust, always verify," requiring verification for every access request based on identity confirmation. Zero Trust Architecture has been explained in the figure below:

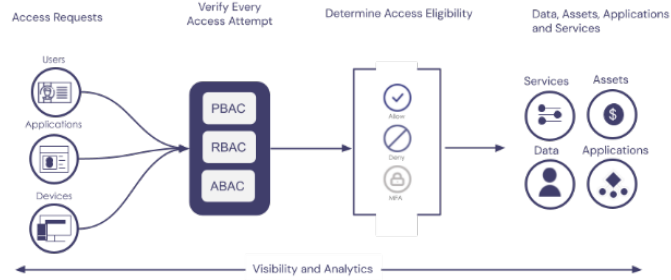

**Figure 1.** Zero Trust Architecture.

According to the ZTA framework, all users, devices, and services requiring access to network resources must first be authenticated and authorized. However, most ZTA implementations rely on static access policies, which pose a significant drawback. These policies often fail to account for the dynamic nature of modern threat environments, including user actions, contextual factors, and time-based risks.

*1.2. The Need for Dynamic Security Solutions*

Static policies, while necessary as a foundational layer, are slow to adapt to anomalies caused by dynamic user activities or environmental changes [2]. For instance, an employee accessing data from a remote location or, an account data or an account exhibiting suspicious activity requesting data can completely nullify static controls and cause breaches. Insider threats further complicate matters, as they originate from legitimate users operating within valid policies.

A more advanced approach, known as dynamic identity-based segmentation, addresses these concerns. By combining real-time behavioral analytics as well as context sensing, ZTA can evolve into a self-improving system. This enables localized threat containment without disrupting legitimate user activities.

One way to formalize dynamic risk assessment is through a risk-scoring model as shown in Equation 1.

$$R = \sum_{i=1}^{n} w_i \cdot f_i(x) \qquad (1)$$

Where:

- $R$: Risk score assigned to a user or device.
- $w_i$: Weight assigned to each behavioral or contextual factor (e.g., login time, geolocation).
- $f_i(x)$: Normalized function representing the observed metric for factor *iii*.
- $n$: Total number of factors considered.

Thresholds ($T$) can then be set to trigger responses:

If $R>T$, then isolate or restrict identity.

*1.3. Research Focus and Objectives*

The research focuses on leveraging AI to enhance ZTA through identity-based threat segmentation. It targets three primary areas to overcome challenges in deploying ZTA solutions. First, behavioral analytics employs AI to scan user activity for logins and resource utilization to create real-time risk scores that enhance threat identification. Second, it integrates with machine learning (ML) models that extend permission based on contextual parameters such as the type of the device being used, time of access, and geographical location to offer better security when managing access. Third, threat segmentation automation plans about how exactly suspicious identities can be quarantined or contained in real-time and how to address the insider threat and motion between networks. These focus suggest providing proactive solutions, contrasting with the static and reactive nature of traditional security measures. Therefore, this research aims to fill key shortcomings of the existing ZTA frameworks to improve cybersecurity while considering the scalability and privacy constraints imperative at large.

*1.4. Real-world Relevance and Use Cases*

The proposed research has immediate practical application in mitigating insider threats within the corporate environment. Another practical implementation scenario pertains to the application of the system in organizations with branches to track and analyze user actions for security threat detection. For example, if an employee accesses a restricted section during off-hours from an unfamiliar device or location, the software flags the behavior as suspicious. By assigning a high-risk score to such behavior, the system can promptly initiate preventive measures, such as temporarily suspending the user's activity and notifying security teams for further investigation. This approach minimizes response time and significantly reduces false alarms, allowing security teams to focus on genuine threats. Further, when the framework extracts all anomalous patterns, all actual threats remain to be served; this guarantees critical resources to focus on appropriate targets. With these real-world applications, the proposed system improves organizational security by accurately assessing risks and implementing timely interventions.

*1.5. Constraints and Challenges*

Implementing the proposed approach involves certain challenges. User data collection and analysis for behavioral modeling raise privacy concerns, requiring robust data anonymization and compliance measures. Moreover, dynamic systems may occasionally misidentify certain legitimate activities as threats, disrupting normal operation [3]. Scalability is another challenge, especially in large and complex networks, as real-time segmentation demands significant computing power.

*1.6. The Impact of Autonomous Threat Segmentation*

Through adopting AI and ML in ZTA, this research seeks to transform organizations' perceptions of network security. The proposed system not only addresses the limitations of static policies but also contributes to strengthening enterprises against insider threats and new-generation cyber attacks.

## 2. Literature Review and Background

*2.1. Zero Trust Architectures: Concepts and Challenges*

Zero Trust Architecture (ZTA) represents a transformative concept in cybersecurity, founded on the fundamental concept of "never trust, always verify" [4]. Unceasing identification and authentication are required to minimize external interference and coaxial control to reduce unauthorized access. ZTA is built on the premise of least privileged network access, a core principle of micro-segmentation that divides the network into isolated zones to contain threats. Some of the key enabling technologies are Multi-Factor Authentication (MFA), Endpoint Detection and Response (EDR), and Software Defined Perimeters (SDP). MFA adds checks to the identity validation processes using at least two factors: EDR provides real-time detection of endpoints, and SDPs create authorized and policy-based perimeters. The following figure shows the concepts related to Zero Trust Architecture:

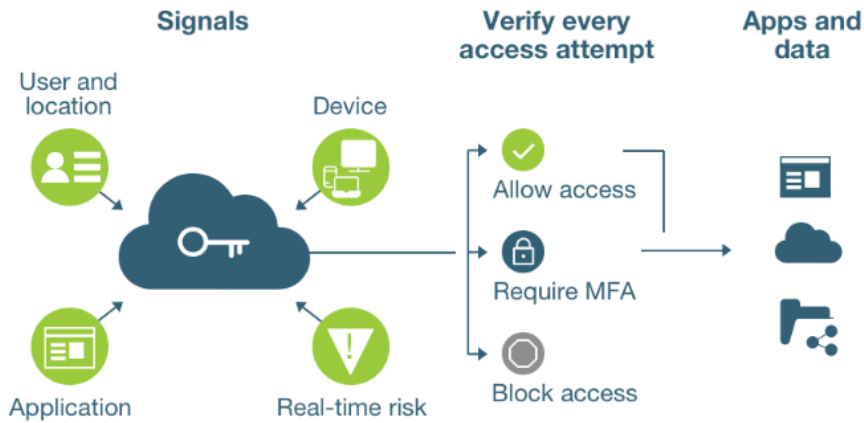

**Figure 2.** Major Concepts Related to Zero Trust Architecture.

Despite its advantages, ZTA faces significant challenges. One weakness is the lack of a dynamic access control model, where permissions that control access to resources are static and fail to adapt to rapidly evolving threats like phishing or malware propagation. Further, ZTA exerts pressure for high resource utilization as a massive infrastructure upgrade and human resources for skillful personnel may also be needed. Syed et al. (2022) also highlight that fixed policies are turning into dynamic strategies that must be relevant enough to tackle the dynamic threat environment and realize the full benefits of ZTA [5]. The following table shows the key technologies in Zero Trust Architecture:

| Key Technologies in ZTA | Role | Challenges |
| --- | --- | --- |
| Multi-Factor Authentication (MFA) | Strengthens identity validation | Increased complexity for users |
| Endpoint Detection and Response | Ensures robust monitoring of endpoints | High resource consumption |
| Software-Defined Perimeters (SDPs) | Establishes secure, policy-driven perimeters | Requires significant implementation effort |

**Table 1.** Key Technologies in Zero Trust Architecture.

*2.2. Behavioral Analytics in Cybersecurity*

Behavioral analytics has emerged as one of the most essential methods in cybersecurity, using AI and machine learning (ML) to identify changes within user behavior trends. This method centers on knowledge of common patterns like login history, access demand, and usage of resources to look for outliers that show signs of malicious activity. For instance, an employee may open files that team members usually do not access at an odd time or from an unfamiliar computer; then an alarm will be raised. Such insights can help an organization prevent the occurrence of insider threats and compromised accounts.

Existing systems like User and Entity Behavior Analytics (UEBA) provide foundational capabilities in this domain [6]. These tools analyze data from logs, network traffic, and other sources to identify unusual behaviors. However, high false positive rates can overload security personnel with non-threatening alerts. Moreover, these systems are designed to operate and may not allow for effective differentiation between minor anomalies and actual threats. For example, if an employee is on a business trip and generates a lot of usage of the geolocation application, they are likely to set off an alert. To overcome these challenges, further developments in behavioral analytics are being made to improve the ML models used in the detection and to incorporate contextual information. The table below provides a comparison of traditional and AI-driven systems:

| Comparison of Behavioral Analytics | Traditional Systems | AI-Driven Systems |
| --- | --- | --- |
| False Positives | High | Lower with advanced ML algorithms |
| Context Awareness | Minimal | Comprehensive with real-time data |
| Adaptability | Static thresholds | Dynamic models |

**Table 2.** Comparison of Traditional and AI-Driven Systems.

*2.3. Context-Aware Access Control*

Context-aware access control is a major leap forward in the cybersecurity approach because it considers several aspects, including the geolocation of the access request, the type of end device used for access, network conditions, and access timing. While conventional systems store fixed forms of the rules or policies that govern the permissions concerning resources, the context-aware do so in real-time. For instance, if a user wants to open certain files, a system may block or decline access if they try to log in from an unknown device or location, even if their account credentials are correct [7].

Machine learning plays a crucial role in enabling these dynamic adjustments. The ability to consult historical data and work in real-time means that the algorithms are equipped for pattern detection and subsequent updating of access policies. The research study by Haque, Bhushan, and Dhiman (2022) reveals that the center of attention, CA systems, enhances the security and use of services by offering varied outcomes based on different situations [8]. However, some issues are still present, including the issue of security as opposed to user accessibility. If proper restrictions are placed, excessive restrictions may become frustrating or find a loophole in the system, making it ineffective. However, incorporating these context-aware solutions calls for high processing power and optimal tuning due to the demand for additional computing time. The following table explores the key features of context-awareness systems:

| Key Features of Context-Aware Systems | Advantages | Challenges |
|---|---|---|
| Dynamic Access Policies | Tailored responses to diverse scenarios | Computationally intensive |
| Real-Time Adjustments | Enhanced security and usability | Risk of introducing latency |
| Variable Evaluation (geolocation, device) | Comprehensive access control | Potential for excessive restrictions |

**Table 3.** Key Features of Context-Aware Systems.

## 2.4. Automated Threat Segmentation

Automated threat segmentation concerns focus on isolating compromised users within a network to limit the spread of threats [9]. It uses techniques of clustering, supervised learning, and anomaly detection to segment the networks dynamically based on threat patterns. For instance, if a device behaves anomalously, it may be promptly moved to a quarantine network section where further analysis is performed.

Threat segmentation efficiency has been proved in one or another research, especially in the contexts of heavy traffic in a network and many endpoints. Cluster analysis puts similar behaviors in one cluster to enable the determination of strange behavior implying a threat [10]. The supervised learning models try to distinguish between different activities based on the training data of classified kernels, thus correctly identifying benign and malicious activities. However, implementing real-time can be problematic since handling the data and calculations requires much computational power. Also, relying heavily on AI-powered systems can lead to the exclusion of some subtle threats.

| Aspect | Traditional Systems | Proposed AI-Driven Systems |
|---|---|---|
| Policy Type | Static | Dynamic |
| Behavioral Analytics | Limited | Advanced with ML models |
| Context Awareness | Minimal | Comprehensive (geolocation, etc.) |
| Scalability | Moderate | High with optimizations |
| False Positives | High | Lower with refined algorithms |

**Table 4.** Comparison Table of Literature.

## 2.5. Zero Trust Architectures and Behavioral Analytics Integration

A research study by Sharma (2021) highlighted the possibility of using ZTA in combination with behavioral analysis to improve cybersecurity [11]. This combination leverages the strengths of both approaches: ZTA's focus on granular access control and behavioral analytics' ability to detect anomalies. For instance, a behavioral analytics engine inside the ZTA architectural model can detect anomalies in user activities and modify access rights in real-time. This integration alleviates the fully realized security policies, which makes the reaction to security threats more dynamic.

It is evident from the research that the integration process can only be seamless when contextual data is allowed [12]. So, access controls can be highly accurate and versatile by enabling definitions based on user roles, device trust levels, and real-time threat intelligence. However, this level of integration is possible only in the context of the most sophisticated ML models, which allow the processing of a wide range of input flows without significant losses in speed or resource utilization.

Another aspect of integrating the concept is its place in a proactive threat prevention narrative. The dynamic modification of controls based on users' activity data and other parameters enables systems to counter threats at an early stage. For example, if the behavioral analytics model predicts existing account tampering, including login sessions in different geographical regions, the system should limit the execution of risky transactions and ask for further identification. The synergistic relationship between ZTA and behavioral analytics helps ensure that threats are contained early, increasing organizational preparedness.

*2.6. Emerging Trends and Future Directions*

The trends in cybersecurity technologies cause improvements in ZTA, behavioral analytics, and threat segmentation. New trends are in realizing data privacy through federated learning, implementing secure access through blockchain, and applying dynamic policy adjustment through reinforcement learning. There is an incentive for privacy since federated learning enables organizations to train ML models together without risking data exposure. Blockchain offers a decentralized yet secure structure for controlling access credentials and the audit trail [13]. Through reinforcement learning, systems can acquire interaction knowledge and improve patterns of access policies in the process.

Further research should be devoted to the computational problems of these technologies. Techniques like edge computing and hardware accelerators could assist in workload partitioning and enhancing real-time responsiveness. Furthermore, it is also important for cybersecurity professionals, data analysts, and company representatives to collaborate and find problems and solutions that fit both business and compliance perspectives [14]. The table below shows the emerging trends in ZTA:

| Emerging Trends | Advantages | Challenges |
| --- | --- | --- |
| Federated Learning | Enhances data privacy | Requires sophisticated infrastructure |
| Blockchain | Decentralized, tamper-proof management | Scalability issues |
| Reinforcement Learning | Dynamic, real-time optimization | Computationally intensive |

**Table 5.** Emerging Trends in ZTA.

## 3. Problem Definition

Zero Trust Architectures face a vital weakness in their reliance on static access policies for managing dynamic and evolving cyber threats [15]. Such policies are highly structured, predefined, and incapable of adapting to variations in user interactions or newly developing security threats in real time. Consequently, organizations struggle to estimate risks and provide apt responses, primarily due to challenges posed by insiders, who comprise a significant portion of today's security threats.

Most current systems lack the ability to analyze dynamic user behavior and generate corresponding risk scores. Without this capability, organizations are unable to detect new threats or anomalies in a timely manner. Furthermore, the absence of adaptive frameworks for access control prevents the consideration of contextual information—such as device trust levels, geographic location, or real-time threat intelligence—further exposing organizational assets to potential exploitation by malicious actors [16].

Additionally, no automation exists to contain the affected identities; thus, the situation worsens. In complex networks, especially those large in scale, once a threat has gained a foothold, the lack of controls to prevent its lateral spread poses the risk of more dangerous breaches being experienced [17]. This makes it easier for individuals to gain access to systems and 'steal' data or disrupt organizational operations.

This paper proposes an autonomous identity-based threat segmentation framework to fill these gaps. With AI-based behavioral analytics combined with Context-Aware Access Control, it is suggested that the framework can improve dynamic risk assessment, real-time policy control, and the prevention of threats. This approach aims to overcome the shortcomings of conventional access policies and enhance the capability of ZTA to protect against diverse and innovative security threats, ensuring better security for organizational capital and resources in the growing challenges posed by novel threats.

## 4. Research Agenda

This study proposes a model that leverages AI to estimate user risk scores based on behavior and enhance authentication within Zero Trust Architectures (ZTA). The primary objective is to develop machine learning (ML) models for context-aware access control policies, enabling dynamic adjustments based on contexts such as device trust, location, and real-time threat intelligence. The second objective is centered on an escalation of automated methods of threat segmentation for early containment of identities that have been infiltrated and containing lateral movements within a network.

To achieve these goals, the methodology will entail capturing login and access data from the enterprise environments to train the models using both supervised and unsupervised learning algorithms. Thus, while designing the system, these objectives will be considered to test the system's true accuracy, false positive rates, and response time on real-world data.

The proposed use case for this research involves testing the system's ability to detect insider threats in a global office environment. Users will be tracked online, access rights will adjust automatically, and as soon as any IDs are infiltrated, they will be immediately quarantined to prevent the breach from expanding. This study aims to offer a specific solution to enhance ZTA's resilience against sophisticated and evolving cyber threats.

## 5. Discussion

*5.1. Behavioral Analytics and Risk Scoring*

One of the key innovations proposed in this research is the use of AI for behavioral analysis. Incorporating the login/access data approach entails reviewing the previous login and access information stripped of identity to identify irregularity in the user's patterns since they indicate the threat. The algorithms are trained on large datasets to identify normal behavior patterns and detect anomalies in real time.

AI Algorithms for Behavioral Analytics

The research suggests using a Random Forest and gradient-boosting algorithm to identify anomalous activities using past patterns [18]. These machine learning models come in handy to detect patterns in an overwhelming majority of user behavior data. Random Forest, which uses several decision trees at once, is proficient in dealing with complex relationships that may be non-linear. Gradient Boosting extends decision trees as it forms them from simpler models with the lowest prediction errors by invoking iterative methods to achieve high levels of anomaly detection.

Additionally, K-mean clustering can be applied to group users with similar behavior. Deviations from these clusters can signify anomalous activity, potentially indicating external threats.

5.1.1. Risk Scoring Formula

Risk scoring is a math concept that was developed to quantitatively define how risky the attempt of a specific user is concerning the access attempt context. Equation 2 shows how to calculate the risk score:

$$\text{Risk Score} = w_1 \cdot A + w_2 \cdot \quad (2)$$

Where:

- *A* represents the anomaly score, quantifying deviations from normal activity, such as unexpected login times or locations.
- *C* represents the contextual score, which evaluates the contextual factors of the access, such as the location, device trust level, and whether the login attempt is from a recognized IP address.
- $w_1$ and $w_2$ are weighting factors determining the relative importance of anomaly and contextual scores in the overall risk assessment.

The anomaly score *A* is derived from the mean value of a defined reference activity and scaled using the Mahalanobis distance, which statistically determines the distance between a point and a distribution. The contextual score *C* can be evaluated according to the context of the access attempt, such as whether a source location or device is secured.

*5.2. Context-Aware Access Control*

Context-aware access control is timely and proactive because it permits access under conditions when specific aspects/parameters filed provide access permission only if it does not exceed a prescribed risk threshold.

5.3. *Dynamic Permission Adjustments*

In a conventional ZTA, the access control is considered blunt, where the user's authorization does not change after the user logs in. However, the proposed system is accompanied by a dynamic model that deals with contextual parameters and makes decisions based on them. For instance, if there is an attempt to connect from a new site or machine, the system may ask for other means of identification, such as MFA, or limit access to critical data until the threat level determination is made.

Another example of context-aware access control is using real-time threat intelligence to modify access permissions [16]. For instance, if an external IP address is recorded as the source of a cyber attack, then the system can block anyone using an IP address of that region or quarantine users connecting from such a region.

*5.4. Automated Threat Segmentation*

The proposed framework is an automated threat segmentation system isolating compromised identities based on access patterns. This segmentation is useful to curb horizontal movement within the network, one of the most common techniques invaders employ once they infiltrate a network.

Graph-Based Models for Threat Segmentation

The segmentation mechanism relies on graph-based models whereby users are modeled as nodes, and access relationships are modeled as connections between those nodes [19]. If a user's access pattern falls in this category, then the system marks that user's node and excludes all his connections with other users or key resources, which isolates the compromised identity. This approach will help stop the threat and its development within a short time.

For example, if the end-user behaves like multiple users from different geographical regions in a short time through login activity, the system alerts or quarantines this user and prevents the user from opening certain forms or other important documents. The user's activity is observed in real-time, and if it persists over time, other containment measures, including network segregation, are initiated.

*5.5. Performance Comparison*

The following table compares the performance metrics of static policies versus the proposed dynamic system:

| Metric | Static Policies | Proposed System |
| --- | --- | --- |
| Response Time | Slow | Real-Time |
| Detection Accuracy | Moderate | High |
| False Positives | High | Low |

1. **Response Time:** Static policies are typically created using manual or time-consuming means of identifying threats, and responses to these threats for the proposed system are achieved instantaneously.
2. **Detection Accuracy:** Static policies may include solutions coded to use simple rules that may not catch new or complex threats. The proposed AI-based system enhances the detection success rates due to the dynamic learning capability of the new behavior patterns.
3. **False Positives:** For any given static policy, many false positives are always possible and rely on complex threats even more. The developed system obtains better model detection and categorization accuracy, although it has some false positives, which are improved during subsequent iterations.

*5.6. Constraints and Challenges*

While the proposed system offers significant improvements over static policies, several constraints and challenges must be addressed:

5.6.1. Privacy Concerns

The collection of user behavior data harbors potential privacy issues as any personal and sensitive information may be collected [20]. To address these issues, the system will first de-identify the data, eliminating any personally identifiable information that might appear in the collected data. Besides, users will be notified of data collection procedures to avoid violating users' data privacy.

5.6.2. Scalability

The scalability of the system is another important consideration. The challenges that come with the growth of organizations and the volume of data require that the system processes large amounts of data in real time. To overcome this challenge, one can use computing resources that are more flexible and can be scaled up or down depending on the data that is generated from a pool of users all over the world.

5.6.3. False Positives

Despite the efforts by the proposed system to minimize false positives, there will be times when legitimate user behavior will be detected as suspicious. To overcome this issue, the models will be iteratively trained, and the findings will be tested, refined, and updated through feedback loops to increase the probability of the predictions being accurate over time.

## 6. Future Trends

Zero Trust Architecture (ZTA) holds great promise, with continued development anticipated, especially concerning modern technologies [21]. These innovations are expected to increase security, flexibility, and effectiveness in response to the increasing sophistication of cyber risks.

*6.1. Federated Learning*

Federated learning is a revolutionary approach to machine learning that uses training models across multiple decentralized data sources without sending the data to a central database. In the context of ZTA, federated learning can enhance behavior analysis and risk-scoring models and safeguard user privacy simultaneously [22]. As data does not need to be sent to a central location, limited information flows through the system with federated learning, protecting data from exposure, especially for entities caught in GDPR compliances. This asynchronous training approach also enhances the practicability of ZTA while guaranteeing data protection when the training occurs.

*6.2. Quantum Computing*

Quantum computing can help transform the capabilities of threat detection and segmentation of the ZTA model. Quantum computing protocols can solve problems relatively faster than classical computers, allowing for real-time analysis of security events [23]. As the computational capability of quantum computing develops, new complex threats can be computed within seconds of their occurrence, and the maximum time an attacker has to act remains minimal. This technology could also enhance complicated cryptographic procedures, forcefully extending ZTA's access control and data security.

*6.3. Blockchain*

Technology such as blockchain can be useful in enhancing the security of processes related to identity verification in ZTA [24]. The identity management system needs a decentralized and immutable ledger that blockchain provides to organizations to support the responsiveness of the information presented by the identity management system. Blockchain enables the access requests and identity transactions to be visible and immutable, thus denying the attackers a chance to mimic or tamper with the authentications. Moreover, the smart contract feature of blockchain may also help implement the access control policies of ZTA in an automated and enforced manner according to the set policies. The following figure shows the integration of ZTA principles with blockchain technology:

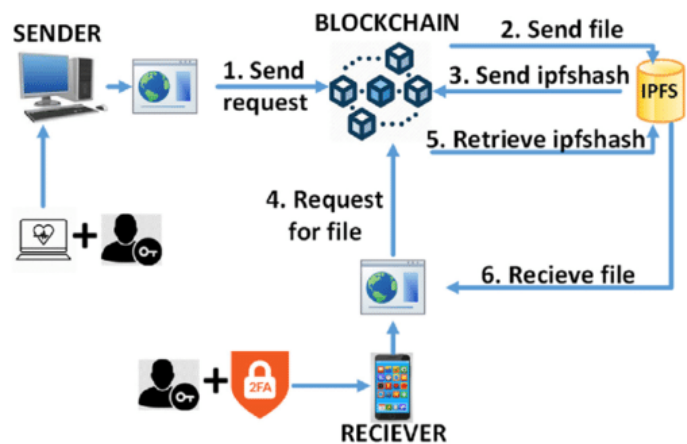

**Figure 3.** Zero Trust Principles with Blockchain.

## 7. Conclusion

This research points out the drawbacks of the conventional identity assurance mechanisms in the traditional Zero Trust Architecture (ZTA) and suggests an advanced AI-augmented approach to dynamically enforce threat segmentation to improve cybersecurity. Traditional ZTA models generally depend on fixed and unchanging access control policies that do not respond to dynamic changes in user activity or new threats. These static systems do not work well when identifying complex attacks, including insider threats prevalent in today's cybersecurity risk landscape. Incorporating behavioral analytics, using context-aware access control, and integrating automated isolation mechanisms in a proposed system provide a better solution to augment ZTA and address dynamic threats of an organization.

Evaluating the AI-driven system in practice, especially in the case of detecting and preventing insider threats, will show that the system can help improve network security while minimizing the impact on productivity. The operations performed in real-time can be time-critical for evaluating all the users and managing the permissions based on collected risk assessment. This dynamic approach is useful in minimizing the threat of unauthorized access while minimizing impairment of bona fide use. However, some difficulties still appear while implementing such a system. One of them is the privacy issue because the system involves identifying and analyzing abnormal user data. This issue has to be solved, and methods like data masking and data protection regulations contact (like GDPR) are helpful in privacy risk diminishment. However, there is the issue of scalability, especially for large organizations with intricate IT systems. Implementing cloud solutions can solve scalability issues due to the ability to access generally available big data resources and guaranteed high performance.

In the future, the use of other sophisticated technologies, such as federated learning and blockchain, to make improvements to ZTA frameworks can be pursued. Another advantage of federated learning is that it makes it easier for the model to train without transferring data to a centralized system, making it more secure. They can provide substantial identity verification and access control updates by offering a decentralized, trustworthy blockchain ledger. When implemented with ZTA, the above technologies will enhance Otto and adaptivity to new threat vectors to improve any organization's security scepter.

## References

1. Admass, W.S.; Munaye, Y.Y.; Diro, A.A. Cyber security: State of the art, challenges and future directions. *Cyber Security and Applications* 2024, *2*, p. 100031.
2. Heartfield, R.; Loukas, G.; Bezemskij, A.; Panaousis, E. Self-configurable cyber-physical intrusion detection for smart homes using reinforcement learning. *IEEE Trans. Inf. Forensics Secur.* 2020, *16*, pp. 1720–1735.
3. Ahmad, A.; Webb, J.; Desouza, K.C.; Boorman, J. Strategically-motivated advanced persistent threat: Definition, process, tactics and a disinformation model of counterattack. *Comput. Secur.* 2019, *86*, pp. 402–418.
4. Roslan, N.I.; Mazman, N.T.; Johari, N.F.A. Zero Trust Architecture: A Paradigm Shift in Network Security. *Authorea Preprints* 2024, p. 56.
5. Syed, N.F.; Shah, S.W.; Shaghaghi, A.; Anwar, A.; Baig, Z.; Doss, R. Zero trust architecture (ZTA): A comprehensive survey. *IEEE Access* 2022, *10*, pp. 57143–57179.
6. Mohanty, R.K.; Kumar, A.P.; Padmaja, R.; Prashanthi, V. Deep Learning for Analyzing User and Entity Behaviors: Techniques and Applications. In *Consumer and Organizational Behavior in the Age of AI*, 2024; pp. 219–250.
7. David, D.S.; Anam, M.; Kaliappan, C.; Selvi, S.; Sharma, D.K.; Dadheech, P.; Sengan, S. Cloud Security Service for Identifying Unauthorized User Behaviour. *Comput. Mater. Contin.* 2022, *70*(2), p. 82.

8. Haque, A.B.; Bhushan, B.; Dhiman, G. Conceptualizing smart city applications: Requirements, architecture, security issues, and emerging trends. *Expert Syst.* 2022, *39*(5), p. e12753.
9. Venne, S.; Clarkson, T.; Bennett, E.; Fischer, G.; Bakker, O.; Callaghan, R. Automated ransomware detection using pattern-entropy segmentation analysis: A novel approach to network security. 2024.
10. Amendola, S.; Spensieri, V.; Biuso, G.S.; Cerutti, R. The relationship between maladaptive personality functioning and problematic technology use in adolescence: A cluster analysis approach. *Scand. J. Psychol.* 2020, *61*(6), pp. 809–818.
11. Sharma, H. Behavioral Analytics and Zero Trust. *Int. J. Comput. Eng. Technol.* 2021, *12*(1), pp. 63–84.
12. Salamkar, M.A.; Allam, K. Data Integration Techniques: Exploring Tools and Methodologies for Harmonizing Data across Diverse Systems and Sources. In *Distributed Learning and Broad Applications in Scientific Research*, 2020; p. 87.
13. Wang, Y.; Li, J.; Yan, Y.; Chen, X.; Yu, F.; Zhao, S.; Feng, K. A semi-centralized blockchain system with multi-chain for auditing communications of Wide Area Protection System. *PLOS ONE* 2021, *16*(1), p. e0245560.
14. Culot, G.; Fattori, F.; Podrecca, M.; Sartor, M. Addressing industry 4.0 cybersecurity challenges. *IEEE Eng. Manag. Rev.* 2019, *47*(3), pp. 79–86.
15. Nahar, N.; Andersson, K.; Schelén, O.; Saguna, S. A Survey on Zero Trust Architecture: Applications and Challenges of 6G Networks. *IEEE Access* 2024, p. 88.
16. Kalaria, R.; Kayes, A.S.M.; Rahayu, W.; Pardede, E.; Salehi Shahraki, A. Adaptive context-aware access control for IoT environments leveraging fog computing. *Int. J. Inf. Secur.* 2024, *23*(4), pp. 3089–3107.
17. Alshamrani, A.; Myneni, S.; Chowdhary, A.; Huang, D. A survey on advanced persistent threats: Techniques, solutions, challenges, and research opportunities. *IEEE Commun. Surv. Tutor.* 2019, *21*(2), pp. 1851–1877.
18. Douiba, M.; Benkirane, S.; Guezzaz, A.; Azrour, M. Anomaly detection model based on gradient boosting and decision tree for IoT environments security. *J. Reliab. Intell. Environ.* 2023, *9*(4), pp. 421–432.
19. Cui, W.; He, X.; Yao, M.; Wang, Z.; Hao, Y.; Li, J.; Cui, W. Knowledge and spatial pyramid distance-based gated graph attention network for remote sensing semantic segmentation. *Remote Sens.* 2021, *13*(7), p. 1312.
20. Nissenbaum, H. Protecting privacy in an information age: The problem of privacy in public. In *The Ethics of Information Technologies*, 2020; pp. 141–178.
21. Alevizos, L.; Ta, V.T.; Hashem Eiza, M. Augmenting zero trust architecture to endpoints using blockchain: A state-of-the-art review. *Security Privacy* 2022, *5*(1), p. e191.
22. Hussain, M.; Pal, S.; Jadidi, Z.; Foo, E.; Kanhere, S. Federated Zero Trust Architecture using Artificial Intelligence. *IEEE Wirel. Commun.* 2024, *31*(2), pp. 30–35.
23. Alauthman, M.; Almomani, A.; Al-Qerem, A.; Al Khaldy, M.A.; Aldweesh, A.; Al Maqousi, A.Y.; Alkasassbeh, M. Quantum computing for cybersecurity: A comparative study of classical and quantum techniques. *Innov. Mod. Cryptogr.* 2024, pp. 75–99.
24. Gupta, A.; Khan, H.U.; Nazir, S.; Shafiq, M.; Shabaz, M. Metaverse security: issues, challenges and a viable ZTA model. *Electronics* 2023, *12*(2), p. 391.